\documentclass[prb,superscriptaddress, preprintnumbers ,twocolumn,showpacs,amsmath,amssymb]{revtex4}
\usepackage{amsmath}
\usepackage{amssymb}
\usepackage{amsthm}
\usepackage{amsfonts}
\usepackage{enumerate}
\usepackage{latexsym}
\usepackage{graphicx}

\newcommand{\be}{\begin{equation}}
\newcommand{\ee}{\end{equation}}
\newcommand{\ba}{\begin{eqnarray}}
\newcommand{\ea}{\end{eqnarray}}
\newcommand{\baa}{\begin{eqnarray*}}
\newcommand{\eaa}{\end{eqnarray*}}

\def\prb#1#2#3{Phys.\ Rev.\ B {\bf #1}, #2 (#3)}

\def\be{\begin{equation}}
\def\ee{\end{equation}}
\def\ba{\begin{eqnarray}}
\def\ea{\end{eqnarray}}

\def\C60{A$_x$C$_{60}$}

\def\HgCu3{HgCa$_2$Cu$_3$O$_{8+y}$}
\def\HgCu4{HgBa$_2$Ca$_3$Cu$_4$O$_{10+y}$}
\def\TlCu{Tl$_2$Ba$_2$CuO$_{6+\delta}$}
\def\TlCu3{Tl$_2$Ba$_2$Ca$_2$Cu$_3$O$_{10+y}$}
\def\TlCu4{Tl$_2$Ba$_2$Ca$_3$Cu$_4$O$_{12+y}$}

\def\BiCu3{Bi$_2$Sr$_2$Ca$_{2}$Cu$_3$O$_y$}

\def\8LSCO{La$_{1.88}$Sr$_{.12}$CuO$_4$}
\def\110LNSCO{La$_{1.5}$Nd$_{0.4}$Sr$_{0.1}$CuO$_{4}$}
\def\stage4LCO{La$_{2}$CuO$_{4+\delta}$}
\def\Y248{YBa$_2$Cu$_4$O$_8$}

\def\NbSe2{NbSe$_2$}
\def\TaSe2{TaSe$_2$}
\def\TiSe2{TiSe$_2$}

\begin{document}
\title{Pairing Symmetry in a Two-Orbital Exchange Coupling Model of Oxypnictides}
\author{Kangjun Seo}
\affiliation{Department of Physics, Purdue University, West Lafayette, Indiana 47907, USA}

\author{B. Andrei Bernevig}
\affiliation{ Department of Physics, University of California, Irvine, CA 92697}
\affiliation{Princeton Center for Theoretical Science, Princeton
University, Princeton, NJ 08544}

\author{Jiangping Hu}
\affiliation{Department of Physics, Purdue University, West Lafayette, Indiana 47907, USA}

\begin{abstract}
We study the pairing symmetry of a two orbital $J_1-J_2$ model for
FeAs layers in oxypnictides. We vary the doping and the value of
$J_1$ and $J_2$ to compare all possible  pairing symmetries in a
mean-field calculation. We show that the mixture of an
intra-orbital unconventional $s_{x^2y^2}\sim \cos(k_x)\cos(k_y)$
pairing symmetry  and a $d_{x^2-y^2}\sim \cos(k_x)-\cos(k_y)$
pairing symmetry is favored in a large part of  $J_1-J_2$ phase
diagram. A pure $ s_{x^2y^2}$ pairing state is favored for
$J_2>>J_1$. The signs of the $d_{x^2-y^2}$ order parameters in two
different orbitals are opposite. While  a small $d_{xy}\sim
\sin(k_x)\sin(k_y)$ inter-orbital pairing order coexists in the
above phases, the intra-orbital $d_{xy}$ pairing symmetry is not
favored even for large values of $J_2$.
\end{abstract}
\date{\today}
\maketitle

High temperature superconductivity (at $56K$) has been recently
reported  in the rare-earth electron and hole-doped oxypnictide
compounds~\cite{Kamihara08, Ren,GFChen1, XHChen, GFChen2, HHWen}.
Preliminary evidence~\cite{Mu,Shan,Liu} suggests that the
superconducting state in the electron-doped oxypnictides, like that
in the Cuprates, has gapless nodal quasiparticle excitations and
hence an unconventional pairing symmetry. A number of
theoretical studies have predicted or conjectured different possible
pairing symmetries, anywhere from $p$-wave to a $\pi$-shifted
$s-wave$~\cite{Dai2008}.

A natural approach to the physics of oxypnictides is by drawing
comparisons with that of the Cuprates.  In the case of Cuprates,
superconductivity is produced by doping a half-filled
antiferromagnetic insulator. The antiferromagnetic exchange
naturally provides for a singlet pairing  amplitude~\cite{Anderson},
with equal mean-field critical temperatures for both $d-wave$ and
extended $s-wave$~\cite{Anderson,Kotliar} in the ultra-Mott limit.
Upon doping, due to the Fermi surface shape, the $d$-wave condensate
has higher mean-field transition temperature than the extended
$s$-wave one~\cite{Kotliar}. The nearest neighbor hoping in the
square lattice of Cu atoms is dominant and provides for a large
Fermi surface with large effective mass in the Mott limit.

The electronic properties of  oxypnictides differ from those in
Cuprates  in several important ways. Most importantly, the undoped
oxypnictides are metallic but their resistivity is strikingly high.
They can hence be interpreted both as a bad metal or as a poor
insulator, leaving open the question of whether a weak or
 strong coupling fixed point governs their physics.  From the band
 structure point of view, barring the existence of un-physically strong
 crystal fields, it seems likely that all $3d$
orbitals of the Fe atoms are involved in the low energy electronic
properties. Numerical results based on first principle calculations
show the presence of small Fermi surfaces~\cite{Singh}. In the unfolded Brillouin
zone consisting of one Fe per unit cell, electron and hole pockets
exist around the $M$ and $\Gamma$, $(\pi,\pi)$ points respectively. The magnetic
properties of the oxypnictides are also different from those of the
Cuprates. Neutron experiments have shown that the magnetic structure
in undoped oxypnictides is not a simple antiferromagnetic
order~\cite{mook} but rather a stripe spin-density wave with  onset
temperature of about $150K$.

The metallic behavior and the existence of Fermi pockets have led to
proposals about the superconducting pairing symmetry and mechanism
which originate in the weak coupling, itinerant limit~\cite{Mazin,Sri}.
In this limit, triplet pairing is possible and has
been predicted~\cite{Xiao}. Numerical and analytic research suggests that the
antiferromagnetic exchange coupling between Fe sites is
strong~\cite{Ma,yildirim,si}. Due to As-mediated hopping, an
antiferromagnetic exchange coupling exists not only between the
nearest neighbor (NN) Fe atom sites but also between next nearest
neighbor (NNN) sites. The NNN coupling strength $J_2$ is comparable
to the NN coupling strength $J_1$. The $J_1-J_2$ model provides for
half-filled magnetic physics consistent with experimental neutron
data~\cite{mook}. A nematic magnetic phase transition has been
predicted in this model~\cite{Fang2008,Xu2008}, consistent with the
experimental observation of a structural transition preceding the
spin density wave formation. Therefore, the magnetic structure of
the undoped oxypnictides suggests that the material is not far from
the strongly coupling, Mott limit.



In the present paper we obtain the superconducting mean-field phase
diagram  of a $t-J_1-J_2$ model with the correct Fermi surface for
the oxypnictide compounds. We predict that two kinds of
intra-orbital pairing order parameters, an extended $s$-wave of the
unconventional form $s_{x^2y^2}\sim \cos k_x \cos k_y$ or a
$d_{x^2-y^2}\sim \cos k_x - \cos k_y$ wave order parameter are the
only possibilities in the Mott limit. A mixture of the
intra-orbital unconventional $s_{x^2y^2}$
pairing symmetry  and the $d_{x^2-y^2}$
pairing symmetry is favored in a large part of  $J_1-J_2$ phase
diagram.  While a small $d_{xy}\sim \sin(k_x)\sin(k_y)$
inter-orbital pairing order coexists in the above phase, the
intra-orbital $d_{xy}$ pairing symmetry is not favored even for
large value of $J_2$ in contradiction with the predictions of
several papers~\cite{si,Li,Xu2008} that rest on an analogy with the
physics in cuprates. While $d_{xy}$ pairing would indeed be favored
for $J_2 \gg J_1$ in the case of a large, single band, Fermi surface
(as in the cuprates)\cite{Subir}, our calculation shows that, for
the oxypnictides' Fermi surface, the $d_{xy}$ order parameter is the
least favored. If the Fermi surface picture emerging from LDA
calculations is correct, we can argue, on general grounds, that
$d_{xy}$ pairing cannot compete with the extended $s$-wave we
propose even if $J_2$ is very large. If we consider a single band
and treat a NNN $J_2$ in mean-field decoupling, the superconducting
transition temperature $T_c$ is self-consistently determined by a
Eliashberg equation $2T_c = J_2 \sum_k(f(k))^2 g(x(k,T_c))$ where
$f(k)$ is the pairing symmetry factor and $g(x)=\frac{tanh(x)}{x}$
(with $x(k,T_c)=\frac{\epsilon(k)-\mu}{2T_c}$) positive and peaked
at the different Fermi surfaces. Hence $T_c$ follows the maximum
value of the pairing symmetry factor $|f(k)|$ close to the Fermi
surfaces. Since in the unfolded Brillouin zone of oxypnictides the
electron pockets are located around $(0,\pi)$, $(\pi,0)$ and the
hole pockets are located around $(0,0)$, $(\pi,\pi)$, for the
$d_{xy}$ symmetry pairing, the pairing symmetry factor, $\sin k_x
\sin k_y$, is always small and $d_{xy}$ pairing symmetry is not
favored even for large $J_2$.
\begin{figure}[t]
\includegraphics[width=7cm, height=5cm]{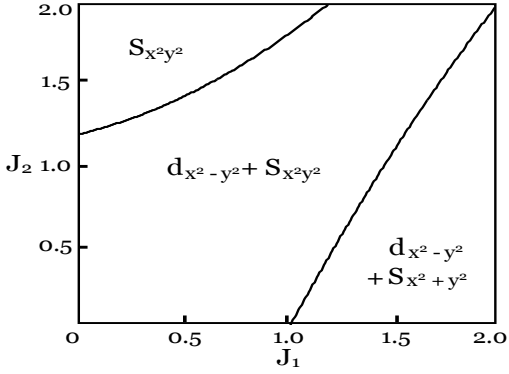}
\caption{The phase diagram in $J_1-J_2$ plane at 18\% electron
doping. } \label{phase1}
\end{figure}

{\it The model:} We focus on a two-orbital per site model of the
oxypnictides, with hybridization between the $d_{xz}$ and $d_{yz}$
orbitals. Although this description is only truly valid in the case
of un-physically large crystal field splitting, we particularize
to this model for analytic simplicity.
We adopt the band structure proposed in Ref.~\cite{Sri}, which at
first sight captures the essence the DFT results:
 \begin{eqnarray}
&H_0 =
\sum_{k\sigma}\psi^\dagger_{k\sigma} T(k) \psi_{k\sigma} \nonumber \\ &T(k)=\left(\begin{array}{cc}
 \epsilon_x(k)-\mu & \epsilon_{xy}(k)\\
 \epsilon_{xy}(k) & \epsilon_y(k)-\mu
 \end{array}\right)
\end{eqnarray}
\noindent where $\psi_{k,\sigma}^\dagger =
(c^\dagger_{d_{xz},k,\sigma}, c^\dagger_{d_{yz},k,\sigma}) $ is the
creation operator for spin $\sigma$ electrons in the two orbitals
and the kinetic terms read: \ba
 &\epsilon_x(k) =  -2t_1\cos k_x - 2t_2\cos k_y -4t_3 \cos k_x \cos k_y \nonumber \\
 &\epsilon_y(k) = -2t_2\cos k_x - 2t_1\cos k_y -4t_3 \cos k_x \cos k_y \nonumber  \\
 &\epsilon_{xy}(k) = -4t_4\sin k_x \sin k_y \label{dispersion1}
\ea \noindent The hoppings have roughly the same magnitude with $t_1
= -1.0, t_2=1.3, t_3=-0.85,$ and $t_4 = -0.85$.  We find that the
half-filled, two electrons per site configuration is achieved when
$\mu=1.54$. The interaction Hamiltonian contains three terms: the
first two are an anti-ferromagnetic NN and NNN coupling between the
spin of identical and opposite orbitals:
 \be H_{i} = \sum_{r\alpha\beta}
J_{i\alpha\beta} (\vec{S}(r,\alpha) \cdot
\vec{S}(r+\delta_i,\beta)-n(r,\alpha) n(r+\delta_i,\beta)) \ee
\noindent
 where $\vec S(r,\alpha)=c^\dagger_{\alpha,r,\gamma}\vec \sigma_{\gamma\gamma'}c_{\alpha,r,\gamma'}$ is the local spin operator,
 $n(r,\alpha)$ is the local density operator, $\alpha, \beta$ are orbital index, $i=1,2$,
$\delta_1$ is the nearest neighbor and $\delta_2$ is the next
nearest neighbor. The third is a Hund's rule coupling of the spins
on different orbitals, on the same site:
\begin{equation}
H_3= -  \sum_{r \alpha} J_H \vec{S}(r,\alpha)
\vec{S}(r,\bar{\alpha})
\end{equation}
\noindent where $\bar{\alpha}$ is the orbital complementary to $\alpha$. This Hamiltonian can be justified through different means.
  Hund's rule is known to play an important effect in Fe, but is usually neglected in the recent calculations on superconductivity in oxypnictides.
  The antiferromagnetic $J_1$ and $J_2$ (both positive) are usually obtained from numerical calculations, although, in the Mott limit,
 they can be justified through a Hubbard-$U$ Gutzwiller method. In the present case, numerical calculations predict a Hubbard $U$ for electrons on the
 same site and same orbital a factor of $2$ larger than the Hubbard $U'$  repulsion of electrons on the same site but different
 orbitals. In this case, the double occupancy constraint can be imposed by a product of the Gutzwiller projectors for the two
 orbitals $P_G = \prod_i (1- n_{i, d_{xz}, \uparrow}n_{i, d_{xz}, \downarrow} )(1- n_{i, d_{yz}, \uparrow}n_{i, d_{yz}, \downarrow} )$. Simple perturbation
 theory leads us to Anderson exchange $\sum_{i,j} 4 (t_{ij}^{\alpha \beta})^2/(U+ 2 J_H) (\vec{S}_{i,\alpha} \vec{S}_{j,\beta} - n_{i,\alpha} n_{i,\beta})
 $ where $\alpha, \beta$ are the two $d_{xz}$ and $d_{yz}$ orbitals. The hoppings in Eq[\ref{dispersion1}], give rise to NN exchange only between spins on
 the same orbitals, and NNN exchange between spins on both the same and opposite orbitals. However, other models for the spin-spin interactions, such as the one in Ref.~\cite{Fang2008},
 which implicitly take into account Hund's rule by formulating the exchange in terms of spin-$2$ or spin-$1$ variables at each site, will also contain NN exchange between spins of
  opposite orbitals. Our mean-field solutions should be interpreted in the same spirit as the superconducting solutions of the
  original $t-J$ model: at some value of the doping, the true undoped spin density wave groundstate disappears and gives way to a superconducting state~\cite{us}.

{\it Mean-field solution in the absence of orbital crossing
exchange:} Keeping all of the above terms becomes analytically
intractable. We proceed with a two-step process: we first mean-field
decouple the interaction Hamiltonian assuming that exchange takes
place only between spins on the same orbitals. While physically
incomplete, this model has the advantage of being analytically
tractable, and exposes the un-competitiveness  of the $d_{xy}$
order. We then numerically solve the full model in a superconducting
mean-field decoupling. The interaction term reads $\sum_{k,k'}
V_{k,k'} c^\dagger_{\alpha,k,\uparrow}
c^\dagger_{\alpha,-k,\downarrow}
                      c_{\alpha,-k',\downarrow} c_{\alpha,k',\uparrow}$ with
 \begin{eqnarray}
& V_{k,k'} = - \frac{2 J_1}{N} \sum_{\pm}  (\cos k_{x} \pm \cos k_{y} )
                               (\cos k'_{x} \pm \cos k'_{y} )  \\
           & - \frac{8 J_2}{N} (\cos k_{x} \cos k_{y}
                               \cos k'_{x} \cos k'_{y} +\sin k_{x} \sin k_{y}
                              \sin k'_{x} \sin k'_{y}) \nonumber
\end{eqnarray}
\noindent with obvious pairing amplitudes in four channels $x^2\pm
y^2$, $xy$ and $x^2 y^2$, $ \Delta_\alpha(k) = \Delta_{x^2+
y^2,\alpha}(k)+\Delta_{x^2-y^2,\alpha}(k) +
\Delta_{x^2y^2,\alpha}(k) + \Delta_{xy,\alpha}(k)$, and
\begin{eqnarray}
&& \frac{\Delta_{x^2 \pm y^2,\alpha}(k)}{\cos k_x \pm \cos k_y}
    = -\frac{2 J_1}{N} \sum_{k'}  (\cos k'_x \pm \cos k'_y ) d(k') \nonumber  \\
&& \frac{\Delta_{x^2  y^2,\alpha}(k)}{\cos k_x  \cos k_y}
    = -\frac{8 J_2}{N} \sum_{k'}  (\cos k'_x  \cos k'_y ) d(k') \nonumber \\
&& \frac{\Delta_{x  y,\alpha}(k)}{\sin k_x  \sin k_y}
= -\frac{8 J_2}{N} \sum_{k'} (\sin k'_x  \sin k'_y ) d(k')
\end{eqnarray}
where $d(k')=\langle c_{\alpha,-k',\downarrow}
c_{\alpha,k',\uparrow} \rangle$.  We use $\alpha =\{1,2\}$ to
represent the orbital index $(xz,yz)$.

We decouple the interaction Hamiltonian with exchange terms only between spins on the same orbitals in mean-field:
$H_m =\sum_k\Psi(k)^\dagger A(k)\Psi(k)$ with
\ba
   A(k)=  \left(\begin{array}{cccc}\epsilon_x(k)-\mu & \Delta_1(k) & \epsilon_{xy}(k) & 0 \\
                               \Delta_1^\ast(k) & -\epsilon_x(k)+\mu & 0 & -\epsilon_{xy}(k) \\
                               \epsilon_{xy}(k) & 0 & \epsilon_{y}(k)-\mu & \Delta_2(k) \\
                               0 & -\epsilon_{xy}(k) & \Delta_2^\ast(k) & -\epsilon_y(k)+\mu
             \end{array}\right)
\ea
with $\Psi(k) = ( c_{1,k,\uparrow}, c_{1,-k,\downarrow}^\dagger, c_{2,k,
\uparrow}, c_{2,-k,\downarrow}^\dagger )$. The particularization to oxypnictides
is present in the hopping terms, which couple different orbitals as in Eq[\ref{dispersion1}].
 $A(k)$ can be diagonalized by an unitary transformatoin, $U(k)^\dagger A(k) U(k)$,
 and the Bogoliubov quasiparticle eigenvalues $ E_1=-E_2$ and $E_3=-E_4$ are given by
 \begin{widetext}
\ba E_{m=1,3}(k) = \frac{1}{\sqrt{2}}\sqrt{(\tilde
\epsilon_x^2+\tilde\epsilon_y^2+2\epsilon_{xy}^2 +
\Delta_1^2+\Delta_2^2)
 \pm \sqrt{(\tilde\epsilon_x^2-\tilde\epsilon_y^2+\Delta_1^2-\Delta_2^2)^2 + 4\epsilon_{xy}^2[(\tilde\epsilon_x + \tilde\epsilon_y)^2+
 (\Delta_1-\Delta_2)^2]}}
\ea
\end{widetext}
where $\tilde\epsilon_{x,y}=\epsilon_{x,y}-\mu$. The self-consistent
gap and density equations are
\begin{eqnarray}
&&\Delta_1(k) =  \sum_{k',m} V_{k,k'} U_{2m}^\ast(k')U_{1m}(k')F(E_m(k')) \\
&&\Delta_2(k) =  \sum_{k',m} V_{k,k'} U_{4m}^\ast(k')U_{3m}(k')F(E_m(k'))\\
&& n_{(1,2)} = 2 \sum_{k',m} U_{(1,3)m}^\ast(k')U_{(1,3)m}(k')F(E_m(k'))
\end{eqnarray}
where $F(E)$ is the
Fermi-Dirac distribution function, $F(E) = \frac{1}{1+e^{E/kT}}$. To
obtain the transition temperature, we linearize the self-consistent
equation for small $\Delta_1$, $\Delta_2$. After tedious
algebra, we find the self-consistent equations around $T_c$
\ba \Delta_2(k) = \sum_{k'} V_{k, k'} (W_3(k')- W_1(k')) \ea where
\begin{equation}\label{w}
W_i = \frac{((\epsilon_x-\mu)^2- \tilde E_i^2) \Delta_2  +
\epsilon_{xy}^2 \Delta_1 }{2|\epsilon_x+\epsilon_y-2\mu|\tilde
E_i \sqrt{4 \epsilon_{xy}^2 + (\epsilon_x- \epsilon_y )^2}}
\tanh(\beta \tilde E_i/2)
\end{equation}
 with $\tilde E_i =E_i(\Delta_1=\Delta_2=0)$.
 \noindent

\begin{figure}[t]
\includegraphics[width=7cm, height=4.5cm]{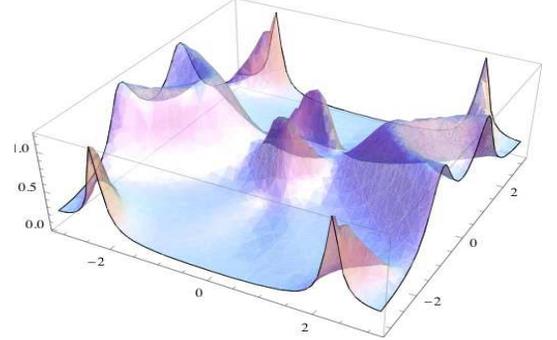}
\caption{A three dimensional plot of the pairing weight $W_3-W_1$ as
a function of $(k_x,k_y)$ (electron doped) by setting
$\Delta_1=\Delta_2=1$ in Eq.\ref{w}.} \label{fig2}
\end{figure}

The above equations can be solved numerically, varying the doping
$\mu$ and the value of $J_1$ and $J_2$. In Fig.\ref{phase1}, we plot
a   phase diagram in $J_1-J_2$ plane with 18\% electron doping. The phase on the left upper corner where
$J_2>J_{2c}\sim 1.2$ has pure extended $s$-wave  pairing  symmetry
$s_{x^2y^2}$ phase.
  The phase on the right  side, where $J_1>J_{1c}\sim1.05$, is a mixture of $d_{x^2-y^2}$
and $s_{x^2+y^2}$. The remaining large part of phase diagram is
described by a phase with mixed $s_{x^2y^2}$ and $d_{x^2-y^2}$
pairing order parameters. This part of phase diagram is believed to
describe the real material since the estimated values of
$J_2\sim0.5$ and $J_1\sim\frac{1}{2}J_2$\cite{Ma,si} are
smaller than $J_{(1,2)c}$. In this mixed state, the  signs of the
$d_{x^2-y^2}$ order parameters in the two orbitals are opposite.
Namely, if $\Delta_1=a \cos(k_x)\cos(k_y)+ b(\cos(k_x)-\cos(k_y))$,
$\Delta_2=a \cos(k_x)\cos(k_y)-b(\cos(k_x)-\cos(k_y))$. Moreover, we
do not find a $d_{xy}$ solution in the entire parameter region.
 Time reversal broken states, such as $s+id$, are not favored either.

 The above results can be understood analytically. First, we can plot
  a pairing weight
$W_3-W_1$  as a function of the Brillouin zone momentum $(k_x,k_y)$
(Fig.\ref{fig2}) by taking $\Delta_2=\Delta_1=1$ in Eq.\ref{w}. The
values of order parameters are determined by the pairing symmetry
factor function times this quantity. The dominant contribution is
clearly around $\Gamma$, $M$ and $(\pi, \pi)$. The $d_{xy}$ order,
in which the pairing symmetry factor, $\sin k_{x} \sin k_{y} $, is
peaked around $(\pm \pi/2, \pm \pi/2)$ has small overlap with the
pairing weight and is not favored. Second, the mixing strength of
two order parameters is determined by multiplying  the two symmetry
factors $(f_1, f_2)$ of two order parameters  and the paring weight: $\sum_{k} f_1(k)f_2(k)(W_3(k)-W_1(k))$. It is easy to check,
for a mixture of $s_{x^2y^2}$ and $d_{x^2-y^2}$, i.e. $f_1= \cos k_x
\cos k_y$, $f_2 =(\cos k_x-\cos k_y)$ that the summation has a large
contribution from the Brillouin zone momentum around the electron
pocket. The mixture strength of the other two order parameters ( $s_{x^2y^2}$ and $
s_{x^2+y^2}$) is very small. This explains why the phase diagram is dominated by
the mixture of $s_{x^2y^2}$ and $d_{x^2-y^2}$. Finally, the difference of the relative sign between
the $s_{x^2y^2}$ and $d_{x^2-y^2}$ order parameters in the two different
orbitals is a result of the fact that exchanging $k_x$ to $k_y$ maps the $xz$ to
the $yz$ orbital.

\begin{figure}[t]
\includegraphics[width=6cm, height=5cm]{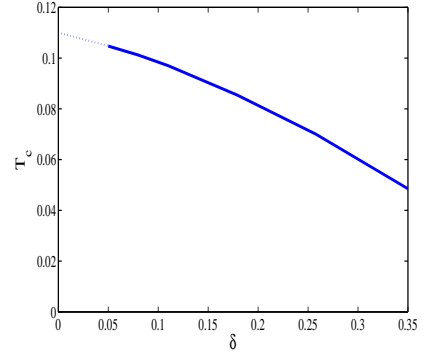}
\caption{The pairing transition temperature as a function of    the
  electron doping concentration at $J_1=0.25$ and $J_2=0.5$. The dashed line indicates the region where the SDW competing phase
takes over from the superconducting phase.
   \label{fig4}}
\end{figure}

The  part of the phase diagram in Fig.\ref{phase1}  with mixed
$s_{x^2y^2}$ and $d_{x^2-y^2}$ pairing becomes larger as the electron doping concentration is reduced: the mixing strength of $s_{x^2y^2}$ and
$d_{x^2-y^2}$ order parameters is (very slightly) increased due to the
enhanced contribution around the $M$ points. In Fig.\ref{fig4}, we plot the
transition temperature as a function of electron doping level at the
fixed values of $J_1=0.25$ and $J_2=0.5$. On the electron-doped
side, $T_c$ is reduced by increasing the doping concentration. This
is similar to Ref.~\cite{Kotliar} and it is, of course, around half
filling, an artifact of the mean-field solution. The true ground
state at half-filling is a spin-density wave~\cite{Dong} which gives
way to a superconductor as the filling is increased.

\begin{figure}[t]
\includegraphics[width=6cm, height=4cm]{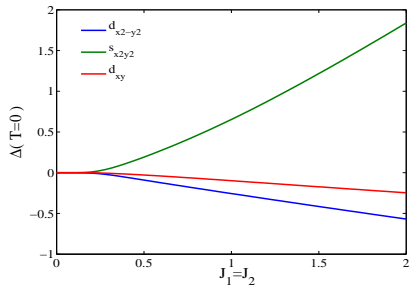}
\caption{The intra-orbital pairing order parameter, $s_{x^2y^2}$,
$d_{x^2-y^2}$ and the inter-orbital pairing order parameter,
$d_{xy}$, as a function of $J=J_1=J_2$ when the chemical potential
is $\mu= 1.8$. \label{fig3}}
\end{figure}

{\it Solutions including orbital crossing exchange coupling and
Hunds coupling:} We now consider the orbital crossing exchange
antiferromagnetic coupling, $J_{1;12}$, $J_{2;12}$ and Hunds
coupling $J_H$. In mean field, we can decouple this interaction in
the particle-particle channel. The orbital crossing exchange
coupling can be decoupled in four spin-singlet orbital crossing
pairing order parameters, $\Delta'(k)=
\Delta'_{x^2+y^2}(k)+\Delta'_{x^2-y^2}(k) + \Delta'_{x^2y^2}(k) +
\Delta'_{xy}(k)$. Hunds coupling can be decoupled to an on-site
spin-triplet, orbital-singlet, order parameter, $\Delta_H=\sum_k
\langle c^{}_{1,k,\uparrow} c^{}_{2,-k,\downarrow}
-c^{}_{2,k,\uparrow} c^{}_{1,-k,\downarrow}\rangle$. The new mean
field Hamiltonian is $H'_m =\sum_k\Psi(k)^\dagger B(k)\Psi(k)$ with
$B(k)= A(k)+\delta A(k)$
\begin{eqnarray}
\delta A(k)=\left(\begin{array}{cccc}0 & 0 & 0 & \Delta'+\Delta_H \\
                               0 &0 & \Delta^{'*} -\Delta^*_H & 0 \\
                               0 & \Delta^{'} -\Delta_H & 0 & 0 \\
                               \Delta^{'*}+\Delta_H^* & 0 & 0 & 0
             \end{array}\right) \nonumber
\end{eqnarray}
We have an additional  self-consistent equation:
\begin{equation}
\Delta'(k)+\Delta_H = \sum_{k',m} V^{'}_{k,k'}
U_{4m}^{'\ast}(k')U^{'}_{1m}(k')F(E^{'}_m(k'))
\end{equation}
where the inter-orbital potential $V'_{k,k'}$ contains NN coupling
$-\frac{2 J_{1; 12}}{N} \sum_{\pm} (\cos k_{x} \pm \cos k_{y} )
(\cos k'_{x}\pm \cos k'_{y} )$, a NNN coupling $ - \frac{8
J_{2;12}}{N}(\cos k_{x} \cos k_{y} \cos k'_{x} \cos k'_{y} +\sin
k_{x} \sin k_{y} \sin k'_{x} \sin k'_{y} )$ and Hund's rule
$-\frac{J_H}{N}$. The self consistent equations are solved
numerically. We find that  in the region where $J_H \sim
Max(J_1,J_2)$, $\Delta_H$  is extremely small. Hence, Hunds coupling
has little effect on pairing symmetry. In the mixed $s_{x^2y^2}$ and
$d_{x^2-y^2}$ phase, for $J_{1;12}\lesssim J_1$ and
$J_{2;12}\lesssim J_2$, the orbital crossing pairing order $\Delta'$
is zero within computing error except for $d_{xy}$.  We find that a
coexisting small inter-orbital paring order with $d_{xy}$ symmetry,
$\Delta'(k)=\Delta'_0 \sin(k_x) \sin(k_y) $. In
Fig.\ref{fig3}, we plot the  result for the intra-orbital pairing
order parameters $s_{x^2y^2}$ and $d_{x^2-y^2}$,  and the
inter-orbital pairing order parameter $d_{xy}$ as a function of
$J=J_1=J_2=J_{1,12}=J_{2,12}$ when the chemical potential is $\mu
=1.8$ - corresponding to $18\%$ electron doping.  The result is a
direct consequence of the $d_{xy}$ symmetry matching between the orbital-crossing pairing and the orbital-crossing hopping term.

{\it Discussion and Summary} Preliminary experimental evidence
suggests the presence of gapless nodal quasiparticle
excitations~\cite{Mu, Shan, Liu} in oxypnictides.   Our model predicts a mixed intra-orbital order parameter $\Delta_{1,2}= \delta_s (\cos k_x \cos
k_y)\pm \delta_d (\cos k_x- \cos k_y)$. In general, the mixed state
is gapped, but with a very small gap. For a
typical value of $\delta_s =0.2$ and $\delta_d =0.1$, the mixed
state will have a gap of $0.025$, which is around one fifth of
the averaged superconducting gap in momentum space. The sign
change of the $d$-wave order parameters between the two orbitals and the
sign change of the $s$-wave order parameter between the hole pockets
and the electron pockets are interesting features of this state. The
sign change can generate new physics, such as new bound
states~\cite{balatsky} formed by impurity scattering.

Although our prediction is based on a two-orbital model, we believe
that the pairing symmetry predicted should be robust even if other
orbitals are added. The pairing symmetry induced by the
antiferromagnetic exchange coupling is mainly determined by the
structure of Fermi surfaces. As the Fermi surfaces in oxypnictides
are located at $\Gamma$ and $M$ points, the $d_{xy}$ paring symmetry
never wins over $s_{x^2y^2}$. Moreover, we find that the
inter-orbital pairing is small even if the orbital crossing exchange
is strong.

{\it Acknowledgments} JPH thanks S. Kivelson, E. W. Carlson, H. Yao
, W. Tsai, C. Fang and D. Yao for  useful discussion. JPH also
thanks X. Tao for important comment and discussion. BAB thanks P. W.
Anderson for useful discussion. JPH and KJS were supported by the
National Science Foundation under grant No. PHY-0603759.

\end{document}